\documentclass[12pt]{article}
\usepackage{latexsym, epsf, proof, lscape}
\usepackage{pxfonts}

\newtheorem{defn}{Definition}[section]
\newtheorem{theorem}[defn]{Theorem}

\newtheorem{lemma}[defn]{Lemma}

\newtheorem{corollary}[defn]{Corollary}
\newtheorem{prop}[defn]{Proposition}
\newcommand{\QED}{\hspace*{\fill}\vrule height6pt width6pt depth0pt}

\newcount\PLv\newcount\PLw\newcount\PLx\newcount\PLy
\newdimen\PLyy\newdimen\PLX \newbox\PLdot \setbox\PLdot\hbox{\tiny.}
\def\scl{.08} 
\def\PLot#1{\PLx`#1\advance\PLx-42\PLy\PLx\PLv\PLx\divide\PLy9%
\PLw\PLy\multiply\PLw9\advance\PLx-\PLw\advance\PLx-4\PLy-\PLy%
\advance\PLy4\PLX=\the\PLx pt\advance\PLyy\the\PLy pt\wd%
\PLdot=\scl\PLX\raise\scl\PLyy\copy\PLdot}
\def\draw#1{\ifx#1\end\let\next=\relax\else\PLot#1%
           \let\next=\draw\fi\next}

\def\IA{\hbox{\PLyy=70pt\draw :::;DMV_gqppyyyyyooooxxxnnwvlutkjaWE%
      =5-./99:::CCCC:::99/..--544=EEWWaajjjkktttttttVVVVVVVV\end
        \hskip7pt}} 
\newbox\IAbox\setbox\IAbox\IA

\begin{document}

\begin{center}\Large
On the Constructive Truth and Falsity \\
in Peano Arithmetic
\end{center}

\normalsize
\begin{center}

Hirohiko Kushida \\

\vspace{1em}
\large
hkushida@gc.cuny.edu\\
Computer Science, Graduate Center\\
City University of New York
\end{center}

\vspace{1.5em}
\small
{\bf Abstract.}
Recently, Artemov \cite{artemov4} 
offered  the notion of constructive
consistency for Peano Arithmetic
and generalized it to constructive truth and 
falsity 
in the spirit of Brouwer-Heyting-Kolmogorov
semantics and its formalization, the Logic of Proofs.
In this paper, 
we provide a complete 
description of constructive truth and falsity for Friedman's constant fragment of Peano Arithmetic.
 For this purpose,
 we generalize the constructive falsity to
$n$-constructive falsity where $n$
is any positive natural number.
We also establish similar classification results for 
constructive truth and $n$-constructive falsity of Friedman's formulas. 
Then, we discuss `extremely' independent sentences
in the sense that they are classically true
but 
neither constructively true nor $n$-constructive false
for any $n$.

\normalsize
\section{Introduction.} 
In the second incompleteness theorem,
G\"odel proved the impossibility to prove
an arithmetical sentence, $Con(PA)=\forall x 
\neg Proof(x, 0=1)$,
which is meant to be a formalization of consistency of Peano Arithmetic,
\textsf{PA}:
 {\it
For all $x$, $x$ is not a code of a proof of $0=1$.}
The formalization 
is concerned with arithmetization of the universal quantifier 
in 
the statement and the arithmetization cannot rule out 
 the interpretability of the quantifier 
to range over both standard and nonstandard numbers.
In a recent paper \cite{artemov4},
Artemov pointed out that it is 
too strong to capture fairly
Hilbert's program on finitary consistency proof
for arithmetic; it asked for a finitary proof that in a formal arithmetic 
{\it no finite sequence of formulas 
is a derivation of a contradiction.  
}
Then, he proposed the notion of constructive consistency,
$CCon(PA)$, 
and
demonstrated that it is actually provable in \textsf{PA}.
%

Moreover, the generalization of constructive consistency 
was offered in \cite{artemov4}
in the spirit of Brouwer-Heyting-Kolmogorov (BHK)
semantics and its formalization, the Logic of Proofs (\textsf{LP}):
constructive falsity with its counterpart,
the constructive truth.
(On 
the family of systems
called Justification Logics
including
the Logic of Proofs, we can refer to
\cite{artemov1, artemov2, af1, af2}.)

\begin{defn}
An arithmetical sentence $A$ is
{\rm constructively false}
if {\rm \textsf{PA}} proves: {\it for any x,
there is a proof 
that `x is not a proof of  A'}.
\end{defn}

This is also viewed as the result of a refinement of
the interpretation of negation and implication 
in the BHK semantics by the framework 
of the Logic of Proofs,
which is compliant with the Kreisel `second clause'
criticism.
(Cf. \cite{af2})

On the other hand, the letterless fragment 
of the logic of provability \textsf{GL}
has been an object of 
modal logical study of Peano Arithmetic, \textsf{PA},
   since Friedman's 35th problem
   in \cite{friedman}.
A letterless sentence is 
one built up from a constant for falsity $\bot$, boolean connectives, and 
the modality $\Box$.
Boolos \cite{boolos1976},
J. van Benthem, C. Bernardi and F. Montagna
 showed that 
 there is a specific normal form
for these sentences and 
the fragment is decidable,
 which was an answer to the Friedman's question.


Following Boolos \cite{boolos},
we call the counterpart of letterless sentences
in \textsf{PA} {\it constant} sentences.
Formally, 
they are built from the sentence $0=1$, a suitable provability predicate
$Prov_{PA}(*)$ and boolean connectives.
Any arithmetical interpretations convert a letterless sentence to the same 
constant sentence in \textsf{PA}.
Here, for the sake of simplicity,
we write $\bot$ to mean $0=1$ 
and $\Box(*)$
 to mean a fixed provability predicate of \textsf{PA}.

In this paper, we are primarily concerned with the constant 
fragment of \textsf{PA};
 in \S 2, we provide a complete delineation of
the constant sentences in terms of the notions of 
constructive tuth and falsity.
Then, it turns out natural to generalize
constructive falsity to
$n$-constructive falsity,
for each positive natural number $n$. 
Also, for each $n$,
we provide
classification results for 
constructive truth and $n$-constructive falsity 
for constant  sentences. 

The `constructive' liar sentence was introduced
and discussed
in \cite{artemov4} along with 
 the Rosser sentence.
In \S 3, we generalize both of these two kinds of
arithmetical sentences,
and  specify the logical status of them  
on the basis of generalized 
constructive falsity.
Also, we clarify which constant sentences
can be the generalized Rosser sentences.

In \S 4, we offer the notion of `extreme' independence from \textsf{PA}
for arithmetical sentences $A$:
both they and their negation are neither 
provable in \textsf{PA} nor belong to 
$n$-constructive falsity for any $n$.
We show that there is an
extremely independent arithmetical sentence
but no constant sentence is extremely independent.




\section{The Constant Fragment of Peano Arithmetic}

In \cite{artemov4},
Artemov clarified the status of some constant sentences
on classical and constructive truth and falsity:
$Con(PA)$ is classically true and 
constructively false.
$0=1$ is classically false and constructively false.
$\neg Con(PA)$ is classically false and 
neither constructively true nor 
constructively  false.
Then, it is natural to ask a general question:
under which condition
a constant sentence is said to be 
constructively true or constructively false.


First of all, we generalize the notion of constructive falsity
to $n$-constructive falsity ($n \geq 1$).
Put $cf^n(F)=\forall x\Box^n\neg(x:F)$
for each $n \geq 1$,
where $\Box^n =\overbrace{\Box\cdots\Box}^n \bot$.

\begin{defn}
An arithmetical sentence $A$ is  $n$-{\it constructive false}
if and only if {\rm \textsf{PA}} proves the sentence $cf^n(A)$.
\end{defn}

The original constructive falsehood is the special case
with $n =1$.

\begin{theorem} (Normal Form Theorem)
$\vdash_{{\rm \textsf{PA}}} cf^n(F) \leftrightarrow. \Box F \to \Box^n \bot. 
$
\end{theorem}

\vspace{1em}
Proof. 
Work in \textsf{PA}.
Suppose $\Box F$, that is, $\exists x(x:F)$ holds. 
Then, for some $y$, we have $y:F$.
By applying $\Sigma_1$-completeness $n$ times, 
we obtain $\Box^n (y:F)$.
On the other hand,
suppose $\forall x \Box^n \neg(x:F)$.
Then, $ \Box^n \neg(y:F)$ holds.
Hence, we obtain $\Box^n\bot$. 
Thus,
$\forall x \Box^n \neg(x:F) \to. \Box F \to \Box^n \bot$. 
For the other direction, obviously, 
$\Box^n \bot \to\Box^n \neg(x:F)$. 
By generalization, 
$\Box^n \bot \to \forall x \Box^n \neg(x:F)$.
On the other hand,
by applying $\Sigma_1$-completeness $n$ times, 
for any $x$,
$\neg(x:F)  \to \Box^n \neg(x:F)$.
By predicate calculus,
$\neg\exists x(x:F)  \to  \forall x \Box^n \neg(x:F)$,
that is,
$\neg\Box F  \to  \forall x \Box^n \neg(x:F)$.
Therefore,
$\neg\Box F \vee \Box^n \bot.  \to  \forall x \Box^n \neg(x:F)$.
\QED

\vspace{1em}
Here we observe some simple facts.

(F1) If $A$ is $n$-constructively false
and \textsf{PA} proves $B\to A$,
$B$ is  also $n$-constructively false.

(F2) If $A$ is $n$-constructively false and $n\leq m$,
$A$ is $m$-constructively false.

(F3)
If \textsf{PA} is $n$-consistent,
that is,
\textsf{PA} does not prove $\Box^n\bot$,
then no $n$-constructively false sentence is
constructively true.

\vspace{1em}
We say that a sentence is $n$-constructively false
{\it at the smallest}
if and only if it is $n$-constructively false but not $m$-constructively false sentence for any $m<n$.

\vspace{1em}
We introduce the following three types of arithmetical sentences.

\vspace{1em}
$(\alpha)$-sentences: of the form $\Box^n\bot \to\Box^m\bot$ $(0\leq n \leq m)$

\vspace{1em}
($\beta, n$)-sentences: of the form $\Box^m\bot \to\Box^{n-1}\bot$ $(1\leq n \leq m)$

\vspace{1em}
$(\gamma, n)$-sentences: of the form: $\Box^{n-1}\bot $ $(1\leq n)$

\begin{lemma}\label{mono}
(1)  $(\beta, n)$- and $(\gamma, n)$-sentences 
are
$n$-constructively false
at the smallest.

(2) ($\alpha$)-sentences are constructively true.
\end{lemma}
 
Proof.  (2) is immediate.
For $(\beta, n)$-sentences, 
consider the formula 
$\Box(\Box^m\bot \to\Box^{n-1}\bot
) \to \Box^k \bot$ with
$0\leq n \leq m$.
This is provably equivalent in \textsf{PA} to
 $\Box^n\bot  \to \Box^k \bot$.
Therefore, \textsf{PA} proves it
if and only if $k\geq n$,
in terms of G\"odelean incompleteness theorems.
The proof is similar for $(\gamma, n)$-sentences.
\QED

\vspace{1em}
By $(\beta\gamma, n)$-sentence we mean 
a conjunction of $(\beta, a)$- and $(\gamma, b)$-sentences
such that $n$ is the minimum of all such $a$'s
 and $b$'s.
In particular, when it consists only of
$(\beta, a)$-sentences,
it is called a $(\beta^+, n)$ sentence.

\begin{lemma}\label{multi}
  $(\beta\gamma, n)$-sentences 
are
$n$-constructively false
at the smallest.

\end{lemma}

Proof.  
Temporarily, let $(\beta, n_i)$ and $(\gamma, m_i) $
denote a $(\beta, n_i)$- and a $(\gamma, m_i)$-sentence,
respectively.
Consider the following sentence.

\begin{center}
\hspace{-7em}
(*)  \hspace{5em}
 $\Box(\bigwedge_i(\beta, n_i) \wedge \bigwedge_j(\gamma, m_j))
\to \Box^k \bot$
\end{center}

where $n = min_{i, j}(n_i, m_j)$.
By using derivability conditions on the provability predicate $\Box$,
this is provably equivalent in \textsf{PA} to the following.

\begin{center}
$\bigwedge_i\Box(\beta, n_i) \wedge \bigwedge_j\Box(\gamma, m_j)
\to \Box^k \bot$.
\end{center}

Furthermore, we can execute the following
transformations, keeping equivalence in \textsf{PA}.

\begin{center}
$\bigwedge_i \Box^{n_i}\bot \wedge \bigwedge_j
\Box^{m_j}\bot
\to \Box^k \bot$;
\end{center}

\begin{center}
$
\Box^{n}\bot \to \Box^k \bot
$.
\end{center}

Thus, in terms of G\"odelean incompleteness theorems,
$(*)$ is provable in \textsf{PA}
if and only if 
 $k \geq n$.
\QED

\begin{lemma} \label{latest}
Any constant sentence is provably in {\rm \textsf{PA}} equivalent to
an $(\alpha)$-sentence or
a $(\beta\gamma, n)$-sentence
for some $n\geq 1$.
\end{lemma}

Proof. 
Boolos' normal form theorem for constant sentences
in \cite{boolos}
states
that  any constant sentence is
equivalent in \textsf{PA}
to a boolean combination of $\Box^n \bot$.
By propositional transformation,
it is further equivalent to
a conjunction of sentences of the form
of $(\alpha), (\beta, n)$ and $(\gamma, m)$.
If it contains only conjuncts which are 
$(\alpha)$-sentences,
it is equivalent to an $(\alpha)$-sentence. 
Suppose that it is of the form $X \wedge Y$
where $X$ contains no $(\alpha)$-sentence
and $Y$ contains only $(\alpha)$-sentences.
As 
$X\wedge Y$ is equivalent in \textsf{PA}
to $X$,
it is a $(\beta\gamma, n)$ sentence with some $n$. 
\QED

\begin{theorem} \label{c1}
Any constant sentence is provably in {\rm\textsf{PA}} equivalent to
 a constructively true sentence or 
an $n$-consistently false sentence for some $n$.
\end{theorem}

Proof.  Derived by Lemmas \ref{multi},  \ref{latest}.
\QED

\begin{theorem} \label{c2}
Let $A$ be any constant sentence
and
$n$ be any positive natural number.
Suppose that {\rm \textsf{PA}} is $n$-consistent.
Then, we have the following.

(1) $A$ is
$n$-constructively false and classically true,
if and only if,
$A$ is provably in {\rm\textsf{PA}} equivalent to
a $(\beta^+, m)$-sentence
for some $m\leq n$. 

(2) $A$ is
$n$-constructively false and classically false,
if and only if,
$A$ is provably in {\rm\textsf{PA}} equivalent to
a $(\gamma, m)$-sentence for some $m\leq n$.

(3) $A$ is constructively true, if and only if,
$A$ is provably in {\rm\textsf{PA}} equivalent to
an $(\alpha)$-sentence.
\end{theorem}

Proof. 
The `if' directions in (1-3) are immediate by
Lemma \ref{multi}.
For the `only if' direction.
(3) is obvious.
We prove (1, 2).
Suppose that $A$ is $m$-constructively false at the smallest 
for some $m\leq n$.
By (F3),
$A$ is not constructively true
and so, is not an $(\alpha)$-sentence.
Since $A$ is constant,
by Lemma \ref{latest},
$A$ is equivalent to 
a $(\beta\gamma, a)$-sentence for some $a\geq 1$.
By Lemma \ref{multi},
$a=m$.

Now, 
if it is classically true,
$A$ is equivalent to 
$(\beta^+, m)$-sentence;
if it is classically false,
$A$ is equivalent to a conjunction of
 $(\gamma, m_i)$-sentences
where $min_{i}(m_i)=m$,
which is equivalent to
a $(\gamma, m)$-sentence,
that is, $\Box^{m-1}\bot$.
\QED

\vspace{1em}

\section{Generalized `Constructive' Liar Sentences
and Rosser Sentences}

In \cite{artemov4},
Artemov offered a constructive version, $L$, of 'Liar Sentence'
by applying the diagonal lemma:

\begin{center}
$\vdash_{\textsf{PA}} L \leftrightarrow 
\forall x \Box \neg (x:L)$\\
\hspace{2.7em}$\leftrightarrow (\Box L \to \Box \bot)$
\end{center}

And he pointed out that $L$ is classically true 
but neither constructively true nor constructively false.
We show that $L$ is $2$-constructively false
and $\neg L$ is ($1$-)constructively false.

We shall introduce a general version of
'Constructive Liar Sentence'.
For each $n \geq 1$, 
$L_n$ is provided by the following.

\begin{center}
$\vdash_{\textsf{PA}} L_n \leftrightarrow 
\forall x \Box^n \neg (x:L_n) $\\
\hspace{2.9em}
$\leftrightarrow (\Box L_n \to \Box^n \bot)$
\end{center}

The existence of $L_n$, we call {\it $n$-constructive liar}, is guaranteed by the diagonal lemma.

\begin{theorem} \label{l1}
(1) $L_n$ is classically true and $(n+1)$-constructively false
at the smallest.

(2) $\neg L_k$ is classically false and $1$-constructively false.
($k\geq 1$)
\end{theorem}

Proof.  For (1). Suppose that $L_n$ is not true.
Then, $\Box L_n$ is not true
and $\Box L_n\to \Box^n\bot$ is true.
This means $L_n$ is true
by definition of $L_n$.
Hence, a contradiction.

Next, again by definition,
\textsf{PA} proves $\Box[L_n
\rightarrow (\Box L_n \to \Box^n \bot)]$
and so 
 $\Box L_n
\rightarrow \Box\Box^n \bot$.
This means $L_n$ is $(n+1)$-constructively false.
To show that $(n+1)$ is the smallest,
suppose that \textsf{PA} proves 
 $\Box L_n
\rightarrow \Box^n \bot$,
 that is,
 $\Box  (\Box L_n \to \Box^n \bot)
\rightarrow \Box^n \bot$.
Then, 
 \textsf{PA} also proves 
 $\Box  \Box^n \bot
\rightarrow \Box^n \bot$,
which is impossible
in terms of G\"odelean incompleteness theorems.

The proof for (2) is similar.
\QED

\vspace{1em}
How about 
G\"odelean Liar Sentence?
It is considered to be $Con(PA)$,
that is, $\neg\Box\bot$.
We can generalize this as follows:
 $n$-{\it G\"odelean Liar Sentence},
 or $n$-{\it Liar Sentence} is defined to be  $Con(PA^n)$,
which is well known to be equivalent to
$\neg \Box^n \bot$.  
\footnote{\textsf{PA}$^n$ is usually 
defined: \textsf{PA}$^0$= \textsf{PA};
\textsf{PA}$^{n+1}= $\textsf{PA}$^n + Con(\textsf{PA}^n)$
    }
About this, we already know its status from the result
of the previous section.
 $Con(PA^n)$ is a $(\beta, 1)$-sentence and,
by Lemma \ref{mono}, it is $1$-constructively false 
at the smallest.
As to $\neg Con(PA^n)$, it is equivalent to $\Box^n\bot$,
which is a $(\gamma, n+1)$-sentence and, by Lemma \ref{mono},
$(n+1)$-constructively false at the smallest.

\vspace{1em}
In \cite{artemov4},
Artemov pointed out that the
Rosser sentence, $R$, is classically true and 
constructively false;
$\neg R$ is classically false and 
constructively false.
Therefore, the result of Rosser's incompleteness 
theorem is said to have been the discovery 
of such a sentence which is
$1$-constructively false and 
the negation of which is also 
$1$-constructively false.

Here again, we can make a generalization:
an arithmetical sentence $R_n$ is an $n$-{\it Rosser sentence}
if both $R_n$ and $\neg R_n$ are $n$-constructively false
at the smallest ($n\geq 1$).
This condition is equivalent to the following:
\textsf{PA} proves

\begin{center}
$
 \neg\Box^k\bot \rightarrow 
(\neg \Box R_n \wedge \neg \Box \neg R_n)$
\end{center}

\noindent
 for any $k\geq n$ and does not for any $k<n$.
The original Rosser sentence $R$
is an instance of $1$-Rosser sentence $R_1$.
It is well-konwn that such an $R_n$ can be constructed in \textsf{PA}.

Now, we can naturally ask: is it possible 
to construct constant $n$-Rosser sentences?

\begin{lemma}\label{r1}
Let $A$ be any constant sentence
containing the provability predicate $\Box$.
If $A$ is $n$-constructively false,
$\neg A$ is $1$-constructively false.
\end{lemma}

Proof.  If $A$ is classically true,
by Theorem \ref{c2},
$\neg A$ is equivalent to the form:
$\bigvee_i
(\Box^{k_i}\bot \wedge\neg \Box^{a_i} \bot)$
where for each $i$, $a_i<n$ and $a_i<k_i$.
Note that in \textsf{PA},
$\bigvee_i
(\Box^{k_i}\bot \wedge\neg \Box^{a_i} \bot)$
implies
$\neg \Box^{min_i(a_i)} \bot$.
We have a derivation in \textsf{PA}:

\begin{center}
\begin{tabular}{lll}$
\Box(\bigvee_i
(\Box^{k_i}\bot \wedge\neg \Box^{a_i} \bot))$ & $\to$ &  
$\Box (\neg \Box^{min_i(a_i)} \bot)$\\
 & $\to$ & $ \Box ( \Box^{min_i(a_i)} \bot\to \bot)$
 \\
 & $\to$ & $ \Box ( \Box\bot\to \bot)$
 \\
  &  $\to$ & $  \Box\bot$
\end{tabular}
\end{center}

If  $A$ is classically false,
by Theorem \ref{c2},
$\neg A$ is equivalent to the form:
$\neg \Box^a \bot$
with $a<n$.
By the hypothesis,
$a\not =0$.
We have a derivation in \textsf{PA}:

\begin{center}
\begin{tabular}{lll}
$\Box\neg \Box^a \bot$ & $\to$ &  
$\Box (\Box^a\bot \to \bot)$\\
  & $\to$ & $ \Box ( \Box\bot\to \bot)$
 \\
  &  $\to$ & $  \Box\bot$
\end{tabular}
\end{center}

Thus, in any case, $\neg A$ is $1$-constructively
false.
\QED

\begin{theorem} \label{r2}
Let $A$ be any constant sentence
containing the provability predicate $\Box$.
Then, the following are equivalent.

(1) $A$ is an $n$-Rosser sentence for some $n$;

(2) $A$ is a $1$-Rosser sentence;

(3) $A$ is $1$-constructively false.
\end{theorem}

Proof. 
Proofs from (2) to (1) and from (2) to (3) are immediate.

From (1) to (2): If (1) holds,
both $A$ and $\neg A$ are both $n$-constructively false
and, by Lemma \ref{r1}, $n=1$.

From (3) to (2): If (3) holds,
 by Lemma \ref{r1}, 
$\neg A$ is $1$-constructively false.
Then, (2) holds.
\QED

\vspace{1em}
By Theorem \ref{r2},
constant sentences can be $n$-Rosser sentences
only when $n=1$.
Of course, we can weaken the definition of $n$-Rosser
sentences:
$R_n$ is a {\it weak $n$-Rosser sentence}
if and only if 
both $R_n$ and $\neg R_n$
are $n$-constructively false (not necessarily at the smallest).

\begin{corollary} \label{r3}
Any constant sentence containing the provability predicate
$\Box$ is a weak $n$-Rosser sentence
for some $n$,
unless it is constructively true.
\end{corollary}

Proof. 
For any constant sentence $A$
containing $\Box$,
if $A$ is not constructively true,
by Theorem \ref{c1},
$A$ is  $n$-constructively false
for some $n\geq 1$.
By Lemma \ref{r1},
$\neg A$ is $1$-constructively, therefore,
$n$-constructively false.
\QED

\vspace{1em}
Also, we obtain a relationship bewtween 
$n$-constructive liar sentences
and $n$-Rosser sentences.

\begin{corollary} \label{r4}
(1) No one of constructive liar sentences 
and the negation of them
is an $n$-Rosser sentence for any $n$.

(2) For any $n\geq 1$,
any $n$-constructive liar sentence $L_n$
is a weak $(n+1)$-Rosser sentence.
\end{corollary}

Proof. 
Derived by Theorem \ref{l1}.
\QED

\vspace{1.5em}
Here is a table to sum up
some of the results from \S\S 2, 3.

\vspace{1.5em}
\begin{tabular}{c|c|c}
    & Classically True & Classically False \\ \hline
    \vdots & \vdots & \vdots \\  \hline
$n$-const. false & $L_{n-1}$, \hspace{.4em} $R_n$
 &  \\ 
         &
$\Box^{m} \bot\to \Box^{n-1} \bot \hspace{.4em}(m\geq n)$ 
 &$\Box^{n-1} \bot$ \\  \hline
\vdots & \vdots & \vdots \\  \hline
$3$-const. false & $L_2$, \hspace{.4em} $R_3$
& \\ 
& 
$\Box^{m} \bot\to \Box^{2} \bot \hspace{.4em} (m\geq 3)$ & $\Box^2 \bot$\\ 
\hline
$2$-const. false & $L_1$, \hspace{.4em} $R_2$
&
\\ 
& 
$\Box^{m} \bot\to \Box^{1} \bot \hspace{.4em} (m\geq 2)$ &  $\Box \bot$\\ 
\hline
 & \hspace{.4em} $R_1$
& 
\hspace{1em} $\neg L_i, \hspace{.4em} \neg R_i  \hspace{.4em} (1\leq i)$
\\ 
$1$-const. false & $\neg\Box^m\bot \hspace{.4em} (m\geq 1)$ & 
$\Box^{m} \bot\wedge \neg \Box^{n-1} \bot 
\hspace{.4em} (m\geq n \geq 2)$ 
 \\ 
& 
& 
$\bot$ \\ \hline
\end{tabular}

\section{`Extremely' Independent Sentences}
We showed that any constant sentence
is $n$-constructively false for some $n$,
unless it is constructively true (Theorem \ref{c1}).
This implies that well-known constant G\"odelean sentences 
such as $Con(PA^n)$ and $\neg Con(PA^n)$ are
$m$-constructively false for some $m$.

How about the `Reflection Principles'?
For any sentence $A$,
let $Ref(A)$ denote  $\Box A\to A$
(what we call the local Reflection Principle
for $A$).
We claim the following.

\begin{theorem} \label{ref1}
For any sentence $A$,
$\neg Ref(A)$ is $2$-constructively false.
\end{theorem}

Proof.  In \textsf{PA}, we have the following derivation.

\begin{center}
$\Box(\Box A \wedge \neg A)$
$\to
\Box\Box A \wedge \Box \neg A$

\hspace{6.2em}$\to
\Box\Box A \wedge \Box\Box \neg A$ \\

\hspace{2.15em} $\to
\Box\Box \bot$
\end{center}

This finishes the proof.
\QED

\vspace{1em}
We note that 
the above argument does not generally hold 
for what we call  the uniform
Reflection Principle.

In addition, it is known that 
$Ref(A^{\Pi_1})$ for $\Pi_1$-sentences $A^{\Pi_1}$
is provably
equivalent to $Con(PA)$ in \textsf{PA}.
Hence, $Ref(A^{\Pi_1})$ is $1$-constructively false
and $\neg Ref(A^{\Pi_1})$ is $2$-constructive false.

These results raise the question of
the status of independence of a kind of G\"odelean sentences
(such as $Con(PA)$, $Ref(A)$, and other constant ones)
from \textsf{PA}.
So, we can naturally ask if there is
a 'truly' independent arithmetical sentence
from \textsf{PA} or not.
We consider stronger notions of
independence.

\begin{defn}
1. An arithmetical sentence $A$ is {\rm strongly independent} from
{\rm \textsf{PA}} if and only if $A$ is neither 
constructively true nor
$n$-constructively false for any $n$.

2. An arithmetical sentence $A$ is {\rm extremely independent} from {\rm \textsf{PA}} if and only if both $A$ and $\neg A$ are strongly independent 
from  {\rm \textsf{PA}}.
\end{defn}

Note that if a sentence $A$ is extremely
independent,
so is  $\neg A$.

\begin{theorem} \label{ref2}
No instance of the local Reflection Principle
is extremely independent from {\rm \textsf{PA}}.
\end{theorem}

Proof. 
Derived by Theorem \ref{ref1}.
\QED

\begin{theorem}\label{limitation}
No arithmetical constant sentence 
 is strongly nor
extremely independent from {\rm \textsf{PA}}.
\end{theorem}

Proof. 
Derived by Theorem \ref{c1}.
\QED

\vspace{1em}
In \cite{artemov4}, Artemov showed that
there is an arithmetical sentence $A$
such that both $A$ and $\neg A$ are not $1$-constructively 
false by using the uniform arithmetical completeness 
for the modal logic \textsf{GL}.
We extend this result to our  general setting.

\begin{prop} (Uniform Arithmetical Completeness for {\rm \textsf{GL}})
There is an arithmetical interpretation $*$
such that for any formula $A$ of modal logic,
$\vdash_\textsf{GL}A$ iff $\vdash_{\textsf{PA}}A^*$.
\end{prop}

This was established independently in \cite{artemov0, avron, boolos1982,
montagna, visser}.

\begin{theorem} \label{extreme}
There is an extremely independent sentence.
\end{theorem}

Proof.  Fix a propositional variable $p$.
It is easily seen that for any positive natural number $n$,
$\not\vdash_\textsf{GL}\Box p \to\Box^n \bot$
and  
$\not\vdash_\textsf{GL}\Box \neg p \to\Box^n \bot$.
(This can be proved by an argument of Kripke completemess or
the arithmetical completeness for \textsf{GL}.)
Therefore, by the above proposition,
there is an arithmetical sentence $F$
such that for any positive natural number $n$, $\not\vdash_\textsf{PA}\Box F \to\Box^n \bot$
and $\not\vdash_\textsf{PA}\Box \neg F \to\Box^n \bot$.
This sentence $F$ is extremely independent from \textsf{PA}.
\QED

\begin{corollary} \label{ref2}
There is an instance of the local Reflection Principle
which is 
strongly independent from \textsf{PA}.
\end{corollary}

Proof.  In the proof of Theorem \ref{extreme},
we obtain the sentence $F$
such that for any positive natural number $n$, 
$\not\vdash_\textsf{PA}\Box F \to\Box^n \bot$.
This sentence is equivalent to
$\Box (\Box F \to F)\to\Box^n \bot$.
Therefore, $Ref(F)=\Box F \to F$ is the desired instance.
\QED

\vspace{1em}
Theorem \ref{limitation}
could signify a limitation of 
the expressibility of
arithmetical constant sentences,
as contrasted with Theorem \ref{extreme}. 

\section{Concluding Remark}
In this paper,
we reported some results on the notion of
constructive truth and falsity in \textsf{PA},
which was just invented 
and has been reported to offer a `real'
solution to the Hilbert program
in Artemov \cite{artemov4}.
In particular,
we showed some theorems 
on the relationship of those notions and 
the `constant' fragment of \textsf{PA},
which has been actively studied a lot since Friedman \cite{friedman}.

As is easily observed,
an arithmetical sentence is $n$-constructively
false 
if and only if 
its unprovability in \textsf{PA}
is provable in \textsf{PA} plus 
$Con(\textsf{PA}^n)$.
As an extension of the work of this paper,
a natural research problem would be
to examine whether or not 
things change 
in an essential way, 
if we are permitted to talk about 
extensions of \textsf{PA} in 
the well-known 
`transfinite progression' since Turing.
Then, we have the notion of $\alpha$-constructive falsity,
where $\alpha$ is an ordinal
in an ordinal system.
As the research subject of the transfinite progression
is known to form a vast area of mathematical logic,
we report this further study in a separate paper.


\vspace{1em}


	
	
\begin{thebibliography}{10{'}}

\bibitem{artemov0} 
S. Artemov, 
Extensions of Arithmetic and Modal Logics (in Russian),
Ph.D. Thesis,
Moscow State University,
Steklov Mathematical Institute, 1979.





\bibitem{artemov1} 
S. Artemov, 
Operational modal logic,
Technical Report MSI 95-29, 
Cornell University, 1995.


\bibitem{artemov2} S. Artemov, 
Explicit provability and constructive semantics,
The Bulletin of Symbolic Logic,
7(1), 2001,  pp.1-36.

\bibitem{artemov4} S. Artemov, 
"The Provability of Consistency",
arXiv preprint arXiv:1902.07404, 2019.


\bibitem{ab} S. Artemov and L. Beklemishev, 
Provability Logic,
In Handbook of Philosophical Logic,
Second edition, Springer, Dordrecht,
pp.189-360.

\bibitem{af1} S. Artemov and M. Fitting, 
Justification Logic,
The Stanford Encyclopedia of Philosophy,
2012.

\bibitem{af2} S. Artemov and M. Fitting, 
Justification Logic: Reasoning with Reasons,
Cambridge University Press,
2019.

\bibitem{avron} A. Avron,
On modal systems having arithmetical interpretations,
Journal of Symbolic Logic,
49, 1984, pp. 935-942.

\bibitem{boolos1976} G. Boolos,
On deciding the truth of certain statements involving the
notion of consistency, 
Journal of Symbolic Logic, 
41, 1976, pp. 779-781.

\bibitem{boolos1982} G. Boolos,
Extremely Undecidable Sentences,
Journal of Symbolic Logic, 
47(1), 1982, pp. 191-196.

\bibitem{boolos} G. Boolos,
The Logic of Provability,
Cambridge University Press,
1993.








\bibitem{friedman} H. Friedman,
  One Hundred and Two Problems in Mathematical Logic,
 Journal of Symbolic Logic, 40:113–129, 1975.







\bibitem{montagna}
F. Montagna,
 On the diagonalizable algebra of Peano arithmetic,
 Bollettino della Unione Mathematica Italiana, 16(5),
 pp. 795-812, 1979.















\bibitem{visser} A. Visser,
 Aspects of Diagonalization and Provability. Ph.D. dissertation, Drukkerij Elinkwijk, 1981.
 
\end{thebibliography}
\end{document}